\documentclass[12pt]{elsart}
\usepackage{epsfig}
 
\begin{document}

\begin{frontmatter}

\title{\it{\textbf{Multifragmentation threshold in $^{93}Nb+^{nat}Mg$\\ 
collisions at 30 MeV/nucleon}}}

\centerline{\large{INDRA and CHIMERA collaborations}}

\author[1]{L.~Manduci}\footnote{Present address : Ecole des Applications Militaires de l'Energie
Atomique, BP 19 50115, Cherbourg Arm\'ees, France},
\author[1]{O.~Lopez},
\author[1]{J.C.~Steckmeyer},
\author[5]{B. Borderie},
\author[1]{R.Bougault},
\author[2]{R.~Dayras},
\author[3]{J.D.~Frankland},
\author[4]{D.Guinet},
\author[4]{P.Lautesse},
\author[5]{N.Le Neindre},
\author[1,6]{M.P\^arlog},
\author[5]{M.F.~Rivet},
\author[7]{E.Rosato},
\author[8]{R.~Roy},
\author[1]{E.~Vient},
\author[7]{M.Vigilante},
\author[1]{B.Tamain},
\author[3]{J.P.Wieleckzo},
\author[9]{S.Aiello},
\author[10]{A.Anzalone},
\author[9]{G.Cardella},
\author[10,11]{S.Cavallaro},
\author[10]{E.De Filippo},
\author[12,13]{S.Femin\'o},
\author[10,11]{F.Giustolisi},
\author[14,15]{P.Guazzoni},
\author[9]{G.Lanzalone},
\author[9]{G.Lanzan\'o},
\author[9,11]{S.Lo Nigro},
\author[9]{A.Pagano},
\author[9]{M.Papa},
\author[9]{S.Pirrone},
\author[9,11]{G.Politi},
\author[10,11]{F.Porto},
\author[10,11]{F.Rizzo},
\author[9,11]{L.Sperduto},
\author[14,15]{L.Zetta}

\address[1]{Laboratoire de Physique Corpusculaire, ENSICAEN, Universit\'e de Caen, CNRS/IN2P3, 
F-14050 Caen Cedex, France}
\vskip -0.15cm
\address[2]{IRFU/SPhN, CEA/Saclay, F-91191 Gif sur Yvette Cedex, France}
\vskip -0.15cm
\address[3]{Grand Acc\'el\'erateur National d'Ions Lourds, 
CEA and CNRS/IN2P3, B.P.~5027, F-14076 Caen Cedex, France}
\vskip -0.15cm
\address[4]{Institut de Physique Nucl\'eaire, CNRS/IN2P3, Universit\'e Claude Bernard Lyon 1, F-69622
Villeurbanne Cedex, France}
\vskip -0.15cm
\address[5]{Institut de Physique Nucl\'eaire,CNRS/IN2P3, Universit\'e Paris-Sud 11,
F-91406 Orsay Cedex, France}
\vskip -0.15cm
\address[6]{National Institute for Physics and Nuclear Engineering, RO-76900 Bucharest-Magurele,
Romania}
\vskip -0.15cm
\address[7]{Dipartimento di Scienze Fisiche and Sezione INFN, Universit\'a di Napoli "Federico II",
I-80126 Napoli, Italy}
\vskip -0.15cm
\address[8]{Laboratoire de Physique Nucl\'eaire, Universit\'e Laval, 
Qu\'ebec, Canada G1K 7P4}
\vskip -0.15cm
\address[9]{INFN, Sezione di Catania, I-95129 Catania, Italy}
\vskip -0.15cm
\address[10]{Laboratori Nazionali del Sud, Universit\'a di Catania, I-95129 Catania, Italy}
\vskip -0.15cm
\address[11]{Dipartimento di Fisica e Astronomia, Universit\'a di Catania,I-95129 Catania, Italy}
\vskip -0.15cm
\address[12]{INFN, Gruppo Coll. di Messina, Messina, Italy}
\vskip -0.15cm
\address[13]{Dipartimento di Fisica, Universit\'a di Messina, Messina, Italy}
\vskip -0.15cm
\address[14]{INFN, Sezione di Milano, Milano, Italy}
\vskip -0.15cm
\address[15]{Dipartimento di Fisica, Universit\'a di Milano, Milano, Italy} 

\newpage

\begin{abstract}
We analyzed the $^{93}Nb$ on $^{nat}Mg$ reaction at $30$ MeV/nucleon in the aim of disentangling
binary sequential decay and multifragmentation decay close to the energy
threshold, i.e. $\simeq 3$ MeV/nucleon. Using the backtracing technique applied to the statistical models 
GEMINI and SMM we reconstruct simulated charge, mass
and excitation energy distributions and compare them to the experimental
ones. We show that data are better described by SMM than by GEMINI in agreement with the fact that
 multifragmentation is
responsible for fragment production at excitation energies around $3$ MeV/nucleon.    
\end{abstract}


\end{frontmatter}

\section{Introduction}

Heavy ion collisions at intermediate energy (20-100 MeV/nucleon) feature several 
mechanisms, such as fusion, 
deep inelastic collisions (DIC) and
direct reactions present at low energy, or incomplete 
fusion and multifragmentation, characteristic of higher energies \cite{fuchs}.
Through heavy-ion collisions, 
very excited nuclei can be produced offering the possibility to study 
nuclei far from their ground state conditions.
The investigation 
of the thermodynamical properties of nuclear matter, such as temperature, 
density, 
excitation energy, can then be carried out. There is intense  
theoretical and experimental research surrounding the debate on the presence
of a liquid-gas phase transition at the origin of the multifragmentation 
process \cite{shlomo}-\cite{richert}. 

At low excitation energies, the decay of hot nuclei follows the predictions
of the statistical model for a compound nucleus, and the deexcitation occurs by binary sequential decays \cite{bethe},\cite{bohr}.
For systems excited at higher energy around 3 MeV/nucleon \cite{bizard}, the
multifragmentation channel opens which is interpreted as a manifestation
of a liquid-gas phase transition \cite{bertsch}.
Many experimental clues have been collected recently \cite{wci},\cite{nato}
on multifragmentation and its relation to phase transition. Nonetheless 
the question is still debated. Disentangling binary
sequential decay from multifragmentation close to the threshold should help to 
better define the thermodynamical properties mentioned above.

In this aim we will study the decay of hot nuclei formed in the 
$^{93}Nb+^{nat}Mg$ reaction at $30$ MeV/nucleon \cite{mandu}.
Using the backtracing technique \cite{des1}-\cite{lope} the data 
will be compared with the predictions of two codes : GEMINI \cite{bob} for binary
sequential decay, and the Statistical Multifragmentation Model (SMM)
\cite{bondorf} for multifragmentation. The reaction dynamics was simulated by Boltzmann-Nordheim-Vlasov (BNV) 
transport model simulations which describe the evolution of the one-body 
distribution function according to the nuclear mean field and including the 
effects of two-body collisions with the test-particle method
\cite{bonasera}, \cite{guarnera}.
We used a soft equation of state ($K=200$ MeV) and a free nucleon-nucleon
collision cross-section $\sigma_{NN}$.
The time evolution of the density contour in the reaction plane 
for various impact parameters was simulated \cite{emma}. Those simulations
\cite{mandu} predict the 
formation of a single source by incomplete fusion for small impact 
parameters ($b\leq 4$ fm).
For larger impact parameters the mechanism showed a binary character
with formation of two sources, quasi-projectile (QP) and quasi-target (QT).
In this paper we will study the decay of incomplete fusion and QP
sources.
%
%
\section{The Experiment}

Collisions of $^{93}Nb$ ions at 30 MeV/nucleon with a $2 mg/cm^2$ thick 
$^{nat}Mg$ target were studied at the GANIL facility using the 
INDRA multidetector. INDRA is constituted by 324 independent telescopes on 
16 rings : the telescopes covering the polar angular range from $3^{\circ}$ to 
$45^{\circ}$ are comprised of an ionization chamber, a 300 $\mu$m thick silicon 
detector and a CsI(Tl) scintillator. 
Those covering the range from $45^{\circ}$ to $176^{\circ}$ are
comprised of two layers : an ionization chamber and a CsI(Tl) crystal
\cite{pouthas1}-\cite{parlog2}.

The reaction $^{93}Nb + ^{nat}Mg$ was measured during a campaign of experiments for which INDRA 
was coupled to the CHIMERA (\cite{aiello} - \cite{pagano2}) first ring, 
which covered the angular range from $1^{\circ}$ to $3^{\circ}$, and some 
INDRA modules in the horizontal plane were replaced
by silicon strip detectors. However, for the present reaction study, none of these
additional detectors were used.
INDRA detects charged particles and fragments with an efficiency 
close to $90\%$ of the whole solid angle and a high 
granularity in order to reduce double counting (down to $5\%$). 
The identification thresholds are low ($\sim 1$ MeV/nucleon). INDRA can
measure ion charge and energy in a wide range and it can resolve masses up to 
Z=4.

The energy calibration of the silicon detectors and ionization chambers for 
heavy fragments was 
obtained by elastic scattering of Ar, Ni and Xe beams on a gold target at 
incident energies from 7 to 9 MeV/nucleon.
For the scintillators the energy calibration was accomplished with secondary
beams of hydrogen and helium isotopes at different energies.
The charge identification was realized by means of a seven parameter fit
of the $\Delta E$-$E$ matrices \cite{tassan} which well reproduces the
form of the lines for each atomic number, $Z$. Unit charge resolution
was obtained for all nuclei produced in this reaction.

The data for the $^{93}Nb+^{nat}Mg$ at 30 MeV/nucleon were recorded with an 
acquisition trigger requiring at least 5 fired telescopes.
The calculated reaction cross section is $\sigma=3.4$ barns \cite{kox}.
The total measured cross-section, calculated from the target thickness,
integrated beam current, and total number of recorded events,
corrected for dead time, is $\sigma=2.$ barns.
The lower experimental value is due to the lack of detection at
forward angles ($< 3^{\circ}$) and, above all, to the acquisition trigger 
condition which eliminates the most peripheral collisions.
For this reaction, the available center of mass energy is $E_{CM}=572$ MeV and the
projectile velocity is $v_{pj}=7.61$ cm/ns. The grazing angle is
$\theta_{graz}=1.44^{\circ}$.

\section{Data Analysis}
\subsection{Event Selection}
\begin{figure}[!h] 
\centering
\epsfig{file=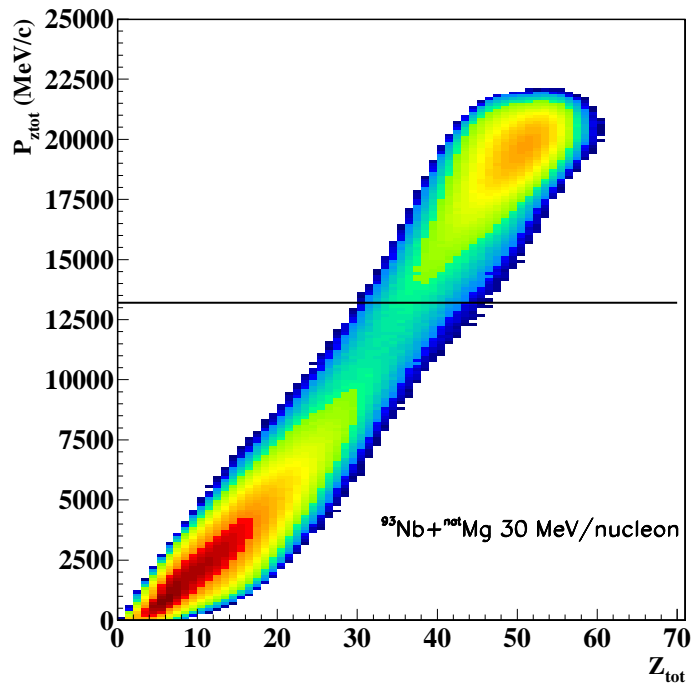,width=10cm}
\caption{Total parallel momentum ($P_{ztot}$) versus 
total charge ($Z_{tot}$) for the $^{93}Nb+^{nat}Mg$ system at 30 MeV/nucleon.}
\label{mom}
\end{figure}
In order to select well-detected events, we used the $P_{ztot}-Z_{tot}$
correlations (see figure~\ref{mom}), the total parallel momentum versus the 
total charge on an event by event basis. 
For the analysis, only the events with a total parallel momentum 
greater than $60\%$ of the 
beam momentum were kept, $P_{ztot}\geq13200$ MeV/c.
\begin{figure}[!h]
\begin{center}
\includegraphics*[scale=0.55]{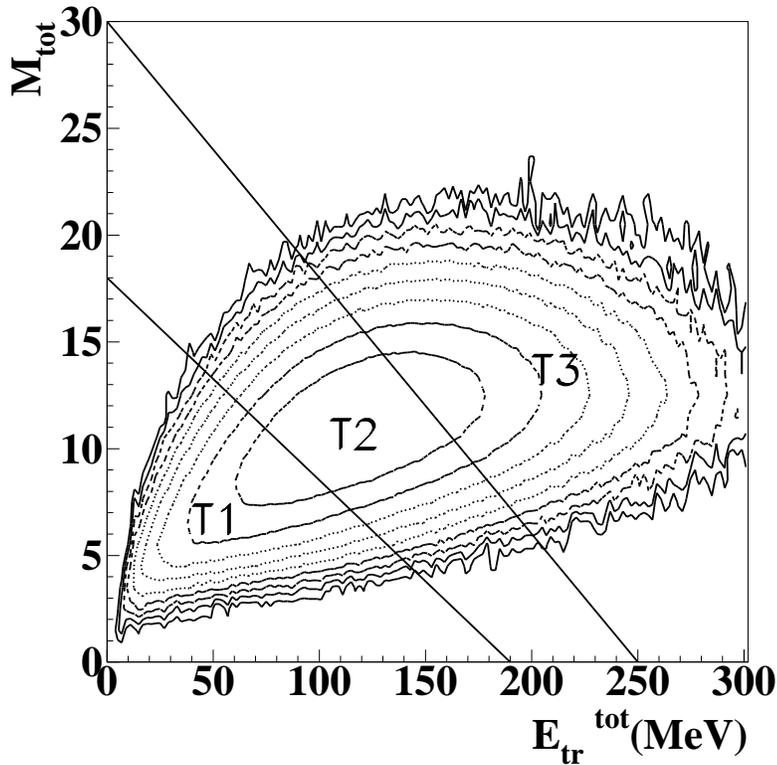}
\caption{\label{molty}Total Multiplicity 
$M_{tot}$ versus Total Transverse Energy $E_{tr}^{tot}$.}
\end{center}
\end{figure}
The events were sorted as a function of two observables correlated with the 
impact parameter \cite{larry}: $E_{tr}^{tot}$, the total transverse energy, 
and $M_{tot}$, the total charged particle multiplicity. 
Three regions are defined according to cuts made perpendicular to the ridge line
of this correlation, which we refer to as T1, T2 and T3 (see figure~\ref{molty}). 
The cuts were made in order to retain approximately the same number of events in 
each region.
Events in the T1 region correspond to less dissipative (peripheral) collisions,
while those in T3 correspond to more dissipative (central) collisions.
Table ~\ref{expsorg} shows the corresponding average values of the 
total transverse energy, the total multiplicity, the residue 
(the biggest fragment) velocity and charge in the three regions.

\begin{table}
\begin{center}
\begin{tabular}{|c|c|c|c|c|}
\hline
 & $<E_{tr}^{tot}>$ $MeV$& $<M_{tot}>$ &$<v_{Res}>$ cm/ns & $<Z_{Res}>$  \\ 
\hline
T1  &$72.9\pm10.1$  & $8.1\pm0.8$ & $6.6\pm0.2$ & $34.1\pm2.3$ \\
\hline
T2&$120.9\pm10.6$  & $11.2\pm0.9$ &$6.4\pm0.2$ & $31.9\pm2.7$  \\
\hline
T3 &$176.4\pm13.8$  & $13.4\pm1.0$ &$6.3\pm0.2$ & $30.5\pm2.8$  \\
\hline
\end{tabular}
\caption{\label{expsorg} Average values of the total transverse energy,
the total multiplicity, the residue velocity and charge for the 3 regions (see text).}
\end{center}
\end{table}

\subsection{Source Reconstruction}

\begin{figure}[!h]
\begin{center}
\includegraphics*[scale=0.8]{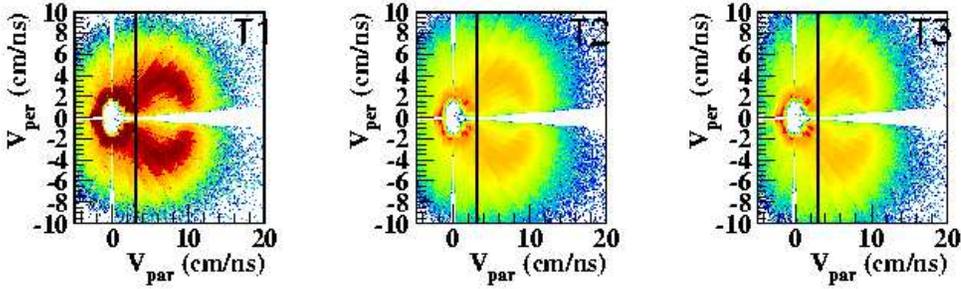}
\caption{\label{vp}Invariant velocity diagrams $V_{per}$ versus $V_{par}$ for
protons of the regions T1, T2 and T3.}
\end{center}
\end{figure}

\begin{figure}[!h]
\begin{center}
\includegraphics*[scale=0.76]{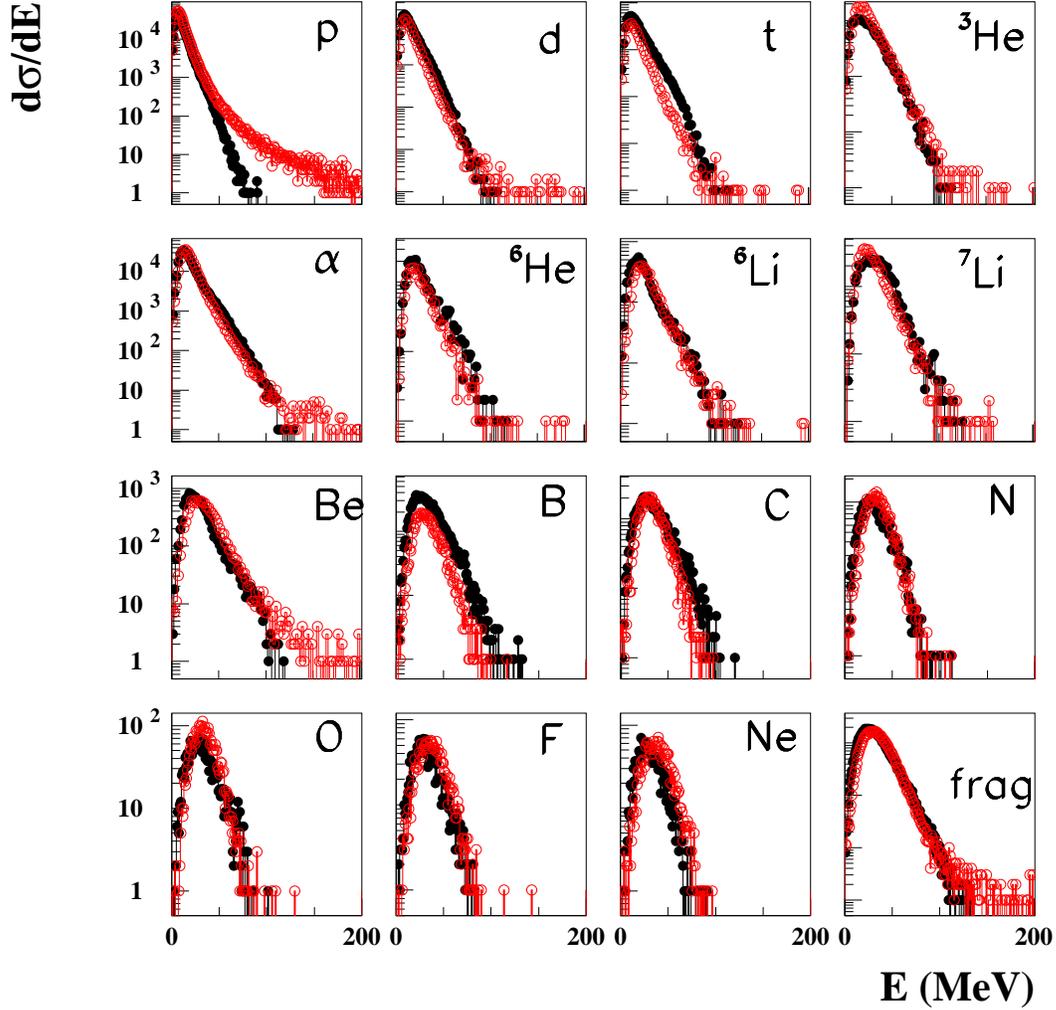}
\caption{\label{t1avi}Backward(black circles)-forward(open circles)
energy spectra in the source reference frame of the region T1 for different nuclear species
ranging from protons to fragments ($Z>10$).}
\end{center}
\end{figure}

%
The source (QP or incomplete fusion) is isolated by a selection in parallel 
velocity of different particles, as shown in figure~\ref{vp}. 
The value of the velocity cut was adjusted, for each region 
and for particles with $Z$ values ranging from $1$ to $10$, in order to 
symmetrize the forward-backward emission in the source reference frame.
Figure~\ref{t1avi} shows the result for the region T1 : the energy spectra of 
the particles emitted in the forward or backward hemisphere of the source 
frame are very similar in most cases. Similar features are observed for the regions T2 and T3. 

Table \ref{tabby} shows the values of the velocity cuts for each
particle for each region.
\begin{table}[!h] 
\begin{center}
\begin{tabular}{|c|c|c|c|c|}
\hline
 Z &A& $v_{lim}^{T1}$ cm/ns  &$v_{lim}^{T2}$ cm/ns & $v_{lim}^{T3}$ cm/ns\\ 
\hline
1 &1  & 3.  & 3. & 3. \\
\hline
1&2 & 3.2 & 3.2 & 3.2  \\
\hline
1&3 & 3.4 & 3.4 & 3.4 \\
\hline
2&3  & 3.8  & 3.8 & 4. \\
\hline
2&4 & 4. & 4. & 4.  \\
\hline
2 &6 & 4. & 4. & 4. \\
\hline
3,4&-  & 4.3  & 4.3 & 4.5 \\
\hline
5,6&- & 4.4 & 4.5 & 4.5  \\
\hline
$>6$ &- & 5. & 5. & 5. \\
\hline
\end{tabular}
\caption{\label{tabby} Velocity cuts for the source selection for the three
regions.} 
\end{center}
\end{table}
The charge $Z_S$ of the source is obtained as the sum of the charge of all 
particles having a parallel velocity $v_{par}$ greater than the cut velocity 
$v_{cut}$.
\begin{equation}
     Z_S = \sum\nolimits_{i=1}^{npart} Z_i  
\end{equation}
Each source velocity component $k=x,y,z$ is evaluated as an average 
weighted by the charge of each particle $i$ of the source :
\begin{equation}
V_s^k = \frac{\sum_{i=1}^{npart}V_i^kz_i}{Z_s}
\end{equation}
Since INDRA does not allow a complete mass determination,
we made the hypothesis that the source has the same isotopic ratio as 
the projectile :
\begin{equation}
     A_{S}=\frac{A_{proj}}{Z_{proj}}*Z_{S}
\end{equation}     
The neutron multiplicity, $m_{neut}$, is estimated as the difference between 
the source mass and the sum of the $A_i$ masses of the different ejectiles :
\begin{equation}
m_{neut} = A_{S} - \sum_i^{npart}A_i
\end{equation}
where the $A_i$ are estimated with the help of two parameterizations : one 
from a cubic polynomial fit on the $\beta$-valley stability :
\begin{equation}
 A_{stable}=1.867Z+0.016Z^2-1.07*10^{-4}Z^3 
\end{equation} 
and the other as in references \cite{bob} and \cite{westfall} :
\begin{equation}
 A_{Charity}=2.08Z+2.9*10^{-3}Z^2 
\end{equation} 
The excitation energy is obtained by calorimetric recontruction :
\begin{equation}
    E^* = \sum\nolimits_{i=1}^{npart} E_i^{kin} + m_{neut}*E_{neut}^{kin}
- \sum\nolimits_{i=1}^{npart}\Delta_i
- m_{neut}*\Delta_n
+ \Delta_{S}\label{ecci}
\end{equation}  
  
In the equation \ref{ecci} $E_i^{kin}$ is the kinetic energy of each particle in the source
reference frame, $\Delta_i$ ($\Delta_S$ for the source) is the mass defect and 
$E_{neut}^{kin}$ is the neutron kinetic energy proportional to the source 
temperature :
\begin{equation}
E_{neut}^{kin} = 2 \alpha T
\end{equation}
where $\alpha = 0.75$ takes into account the temperature decrease as the source 
emits neutrons \cite{steck}.

\begin{table}[!h]
\begin{center}
\begin{tabular}{|c|c|c|c|}
\hline
& $T1$  & $T2$ & $T3$\\
\hline 
$Z_{S}$ & $40.7\pm2.0$&$42.8\pm2.1$ &$45.1\pm2.0$\\
\hline 
$A_{S}$ &$91.2\pm4.7$ &$96.0\pm4.9$ &$101.3\pm4.6$ \\
\hline 
$M_p$ &$2.0\pm0.6$ &$2.7\pm0.7$ &$3.1\pm0.7$ \\
\hline
$M_{\alpha}$ &$1.1\pm0.5$ &$1.8\pm0.6$&$2.5\pm0.7$\\ 
\hline
$M_{frag}$ &$1.2\pm0.4$ & $1.4\pm0.3$ &$1.6\pm0.3$\\
\hline 
$M^{stable}_{neut}$ & $2.8\pm1.1$&$4.8\pm1.3$&$6.4\pm1.4$\\
\hline 
$\epsilon_{stable}^* MeV/A$ & $1.6\pm0.4$&$2.6\pm0.4$ &$3.4\pm0.5$\\ 
\hline 
$M^{Char.}_{neut}$ & $6.1\pm0.6$ & $7.63\pm0.8$ & $8.7\pm1.0$\\ 
\hline 
$\epsilon_{Char.}^* MeV/A$ & $ 2.2\pm0.3$&$3.1\pm0.4$ &$3.8\pm0.4$ \\
\hline
\end{tabular}
\caption{\label{sorg}Characteristics of the reconstructed source : charge, 
mass, light particle, fragment and neutron multiplicities and excitation energy.}
\end{center}
\end{table}

In table ~\ref{sorg} are reported the average characteristics of the reconstructed 
source for the three 
regions : the charge $Z_{S}$, the mass $A_{S}$, the average 
multiplicities for protons $M_p$, alphas $M_{\alpha}$, fragments $M_{frag}$
(biggest one included); the neutron multiplicity $M^{stable}_{neut}$ and 
excitation energy obtained with $A_{stable}$ mass parameterization; the neutron 
multiplicity $M^{Char.}_{neut}$ and the excitation energy obtained with 
$A_{Charity}$ mass parameterization. 
As it can be seen from the table, the two sets of excitation energies differ by
about 0.5 MeV/nucleon. This is mainly due to the difference in the deduced 
neutron 
multiplicity with the two possible choices for the residue mass.
Although the average fragment multiplicity is low, the excitation energy 
for the 
regions T2 and T3 is about $\sim 3$ MeV/nucleon, value at which experimental 
results agree to set the multifragmentation 
threshold \cite{pocho},\cite{bizard}.

\section{GEMINI and SMM Backtracing Simulations}

We now compare the data with the results of two statistical codes, GEMINI \cite{bob} and SMM \cite{bondorf}. 
One expects that experimental results are
relative to an ensemble of various sources (defined by 
{\it{distributions}} of charge, mass and excitation energy).
We therefore applied a backtracing procedure (\cite{des1} -\cite{lope}) to GEMINI and to SMM 
simulations in order to determine these distributions.

\begin{figure}[!h]
\begin{center}
\includegraphics*[scale=0.7]{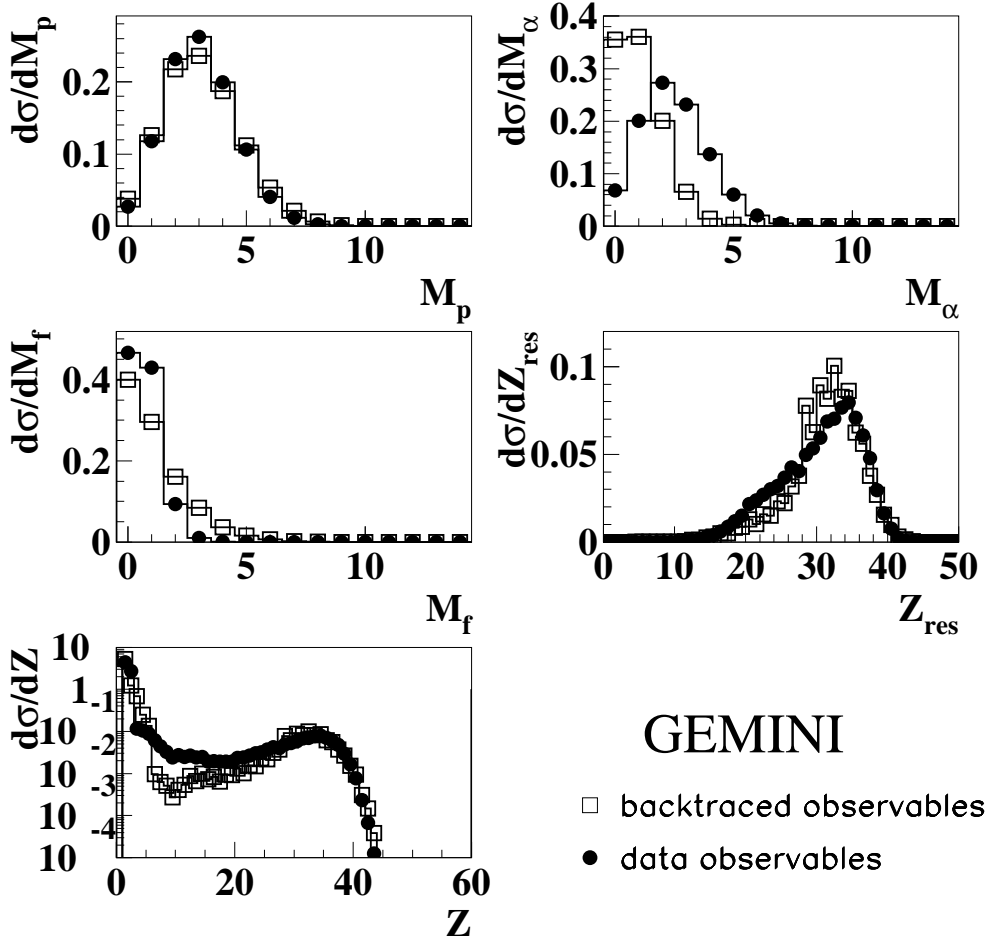}
\caption{\label{gem_obs}Comparison between data (full points) and GEMINI
results (open squares) for region T3. In the top row : $M_p$ : proton
multiplicity and $M_a$ : alpha multiplicity; in the middle row : $M_f$ :
fragment multiplicity and the residue charge distribution. In the bottom row : the
total charge distribution.}
\end{center}
\end{figure}

The procedure begins by first choosing the observables which
will be used to constrain the source distributions.
In our case we chose the multiplicity distributions of protons, alphas and fragments and the charge 
distribution of the residue and of all the products.
This choice was made because these observables do not depend on the hypothesis
taken for the mass of the source.
In a second step, the procedure explores a 4-dimensional space constituted by the source variable parameters : 
charge $Z_0$, mass $A_0$, excitation energy $\epsilon_0$ and angular momentum $L_0$
(for GEMINI) or rotational energy $E^{rot}_0$ (for SMM).
The source parameter distributions are extracted from this 4-dimensional space
with the help of the Kolmogorov-Smirnov test applied to simulated and experimental observable distributions.
The simulated events are filtered in each step using a software replica of the experimental apparatus and 
conditions.
This iterative procedure converges to give the parameter distributions of the simulated sources, 
which can then be compared with those reconstructed experimentally in order to test the agreement 
of data with both models. \par

The region T1, representing the most peripheral events, could be affected by an experimental 
bias due to the trigger condition which suppresses low multiplicity events. Conversely, the region
T3, representing more dissipative events, is supposed to be less altered. Therefore we focused
our attention on T3, this latter being of major interest for the study of multifragmentation 
threshold.  
Figure~\ref{gem_obs} shows the comparison between experimental observables 
(full circles) and backtraced observables from GEMINI (open squares) for region T3.
The charge distribution is poorly reproduced, with a lack of fragment production
 with $Z=6-20$ together with a strong underestimation of the alpha particle
multiplicity. In this case the average angular momentum from the GEMINI backtracing is about 
$30-35$ $\hbar$, sufficiently high to permit complex particle emission. This spin value is in
agreement with other findings on rather similar systems ~\cite{steck2008}. \par
\begin{figure}[!h]
\begin{center}
\includegraphics*[scale=0.50]{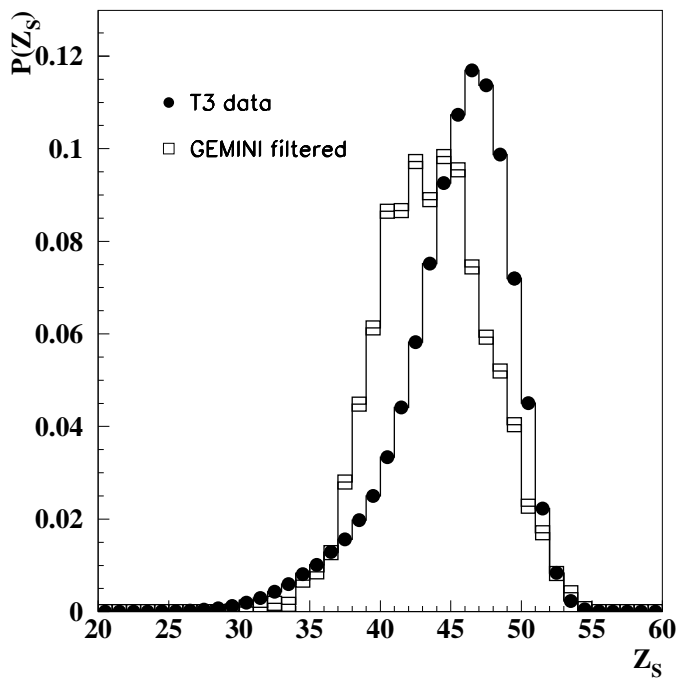}
\caption{\label{gem_var1}Source charge distributions : experimental 
(full points), GEMINI backtraced filtered (open squares) for region T3.}
\end{center}
\end{figure}

The maximum of the source charge distribution, shown in Figure ~\ref{gem_var1},
is shifted to lower $Z$ values as compared to the experimental data,
although both distributions have the same mean value (see Table ~\ref{tab4}).

\begin{figure}[!h]
\includegraphics*[width=7.5cm]{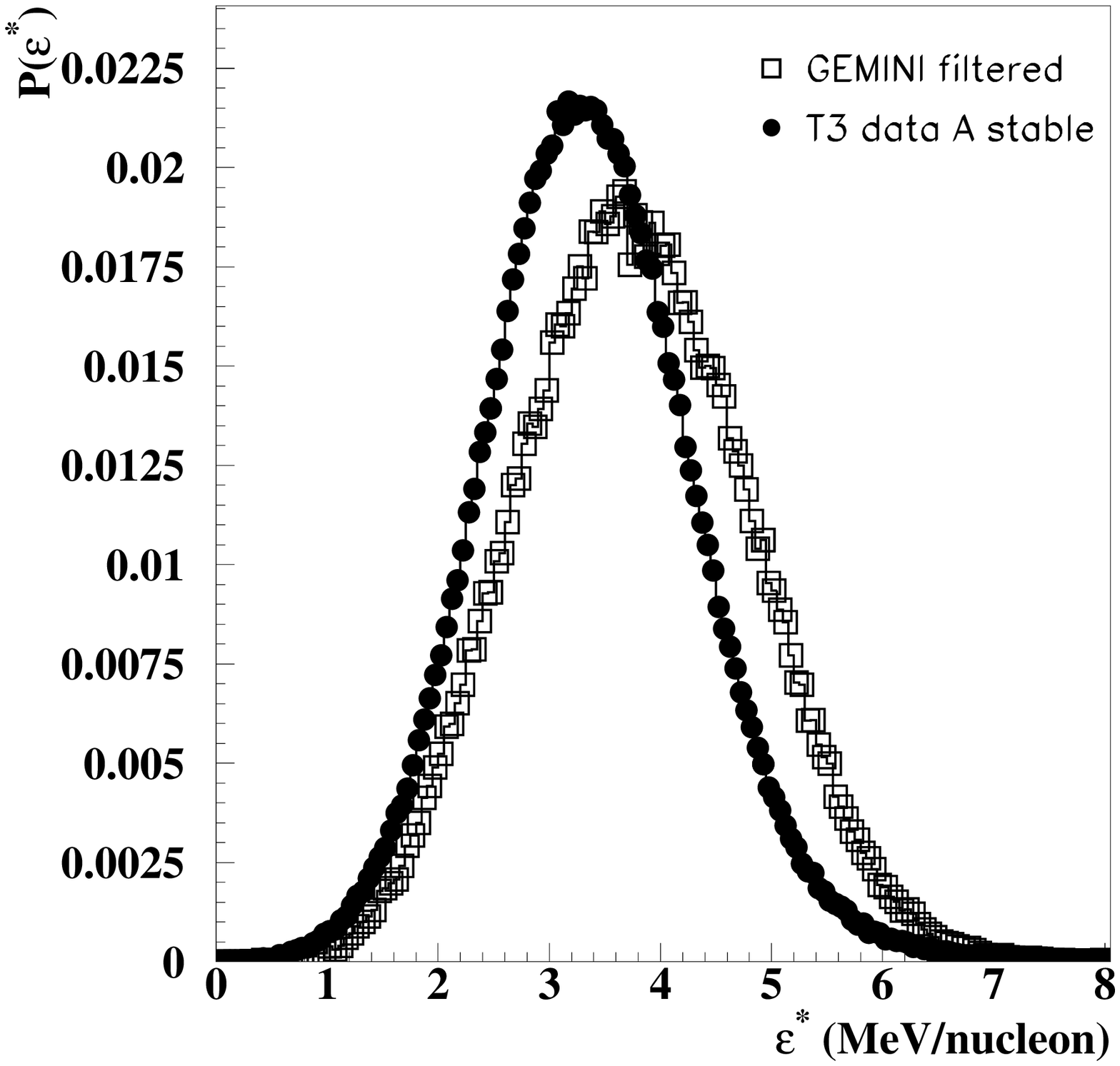}
\includegraphics*[width=7.5cm]{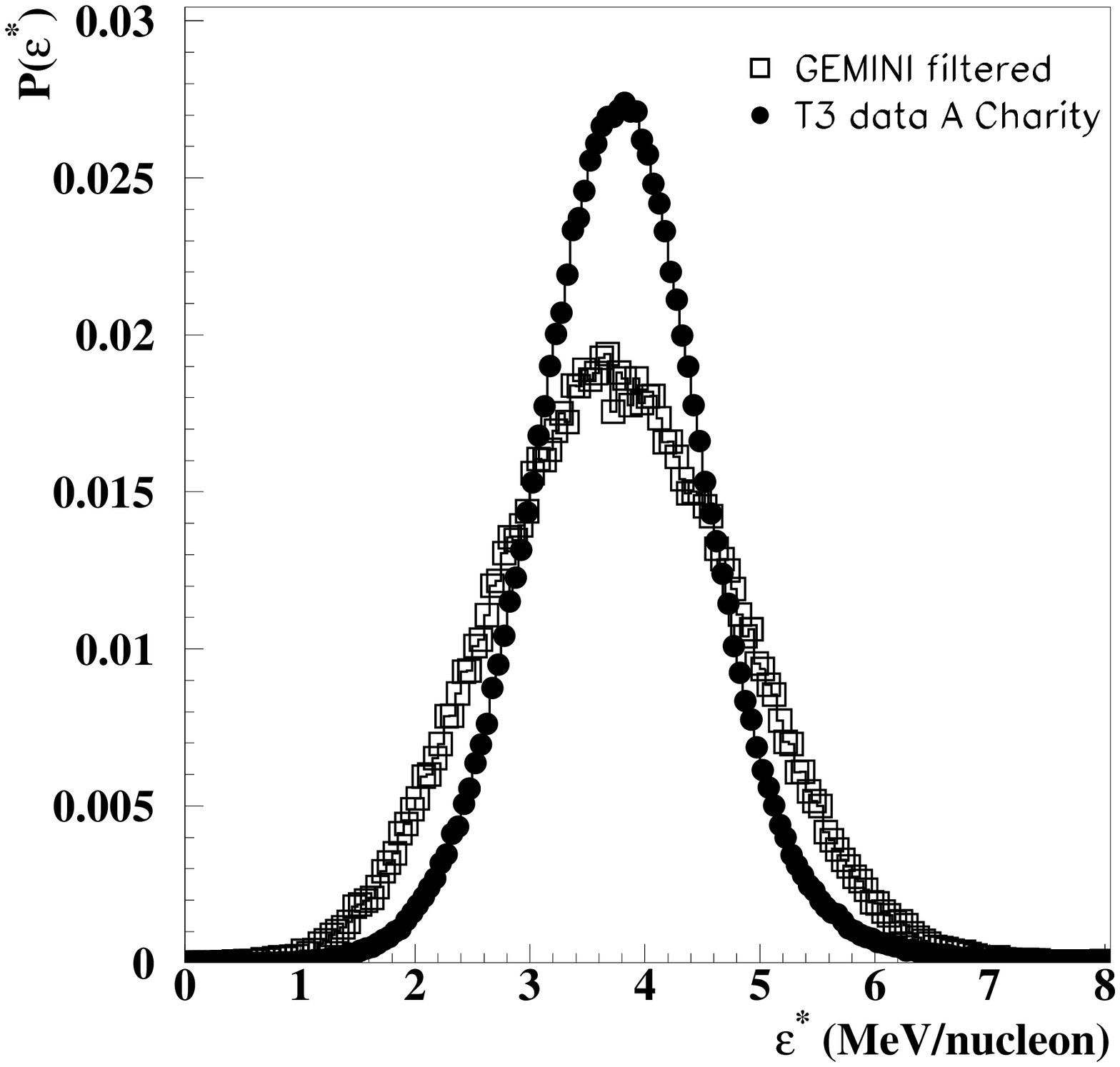}
\caption{\label{conf}Source excitation energy for $A_{stable}$ (left) and $A_{Charity}$
(right) mass parametrization. GEMINI backtraced (open squares) and data 
excitation energy (full points). Region T3.}
\end{figure}
Figure~\ref{conf} displays the excitation energy distributions for $A_{Charity}$ 
and $A_{stable}$ parameterizations. It is possible to note a better agreement 
for the data treated with the $A_{Charity}$ mass 
parameterization since the mean values of the compared distributions are
coincident. The comparison with the data according to the other mass
parameterization shows a shift of about 0.5 MeV/nucleon.

\begin{figure}[!h]
\begin{center}
\includegraphics*[scale=0.7]{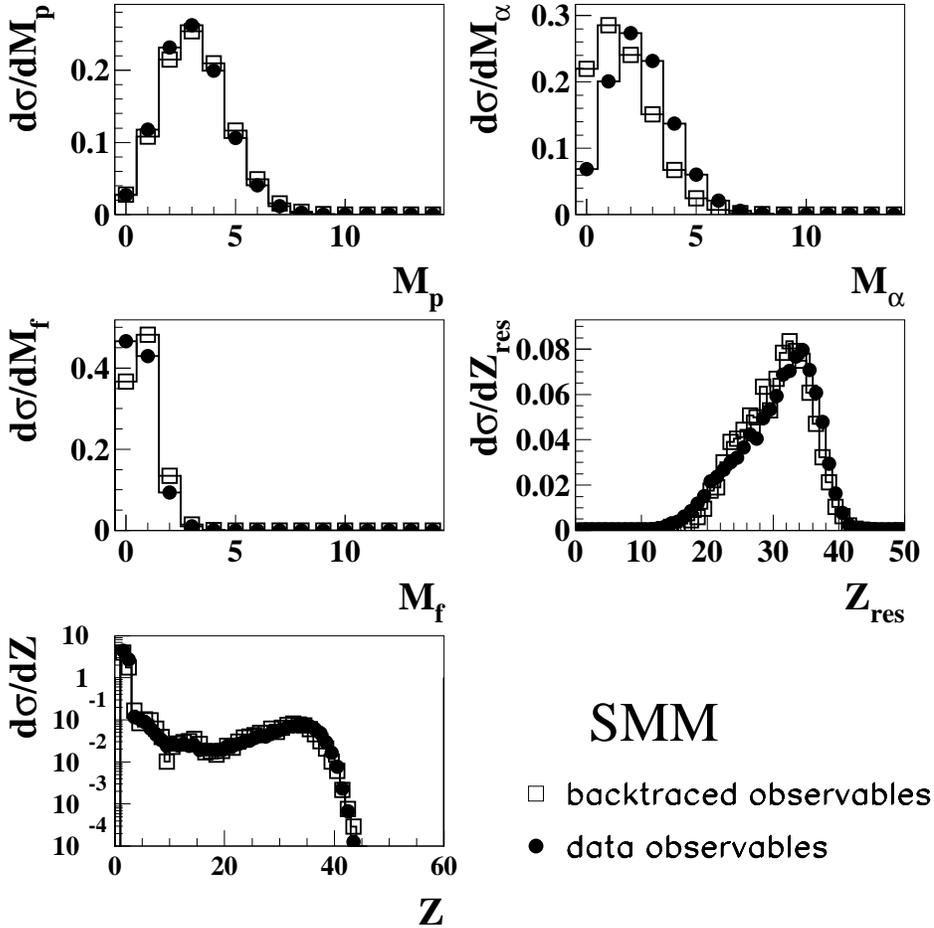}
\caption{\label{SMM_back}Comparison between data (points) and SMM
results (squares) for region T3. In the top row : $M_p$ : proton
multiplicity and $M_a$ : alpha multiplicity; in the middle row : $M_f$ :
fragment multiplicity and the residue charge distribution. In the bottom row: the
total charge distribution.}
\end{center}
\end{figure}

\begin{figure}[!h]
\begin{center}
\includegraphics*[scale=0.50]{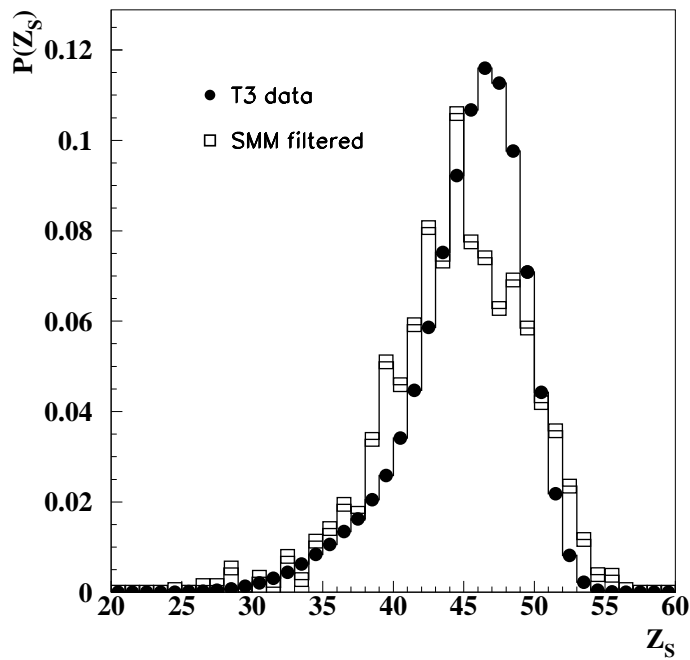}
\caption{\label{SMM_var1}Source charge distributions : experimental
(full points) and SMM backtraced results (open squares) for region T3.}
\end{center}
\end{figure}

\begin{figure}[!h]
\includegraphics*[width=7.5cm]{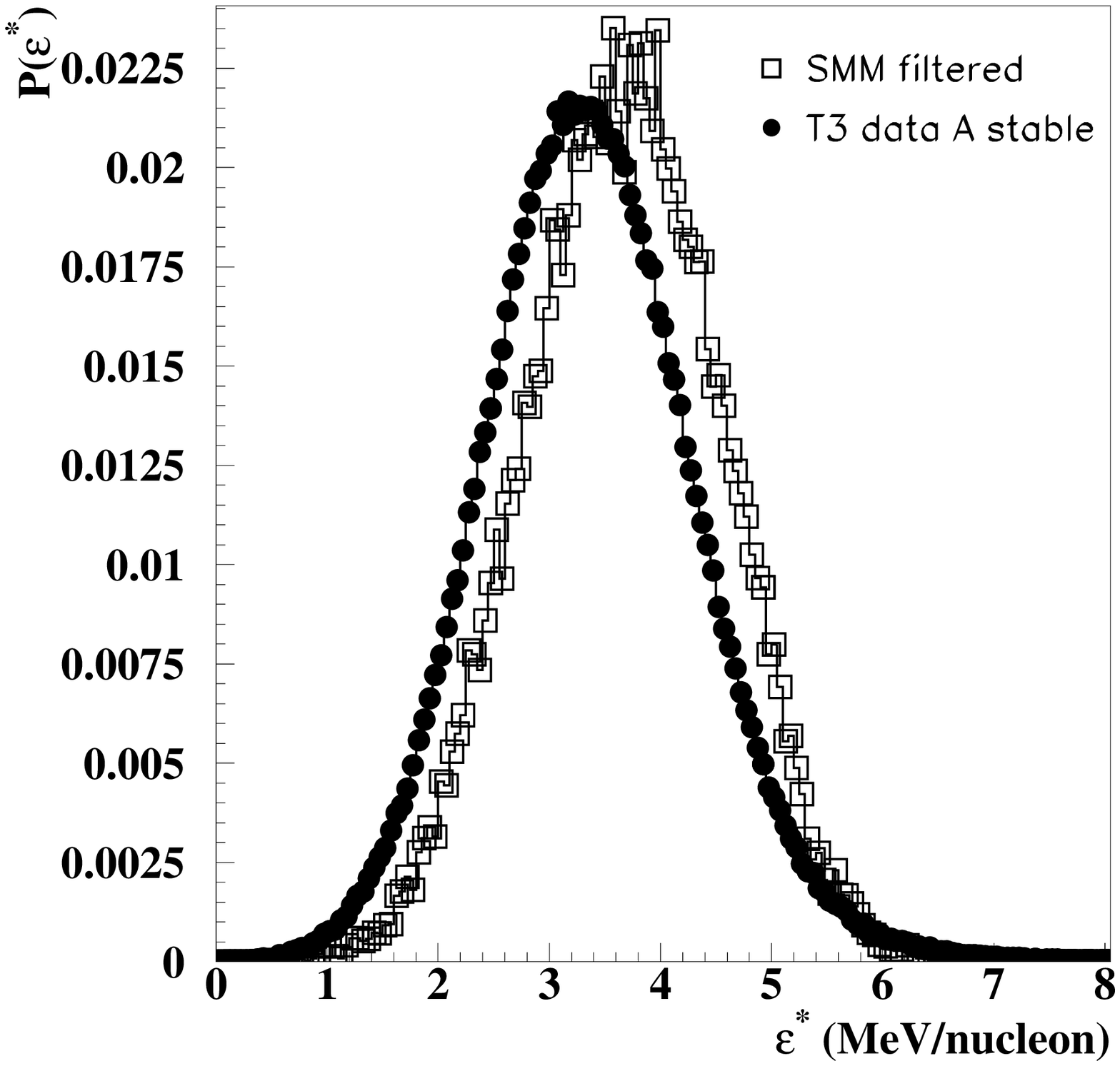}
\includegraphics*[width=7.5cm]{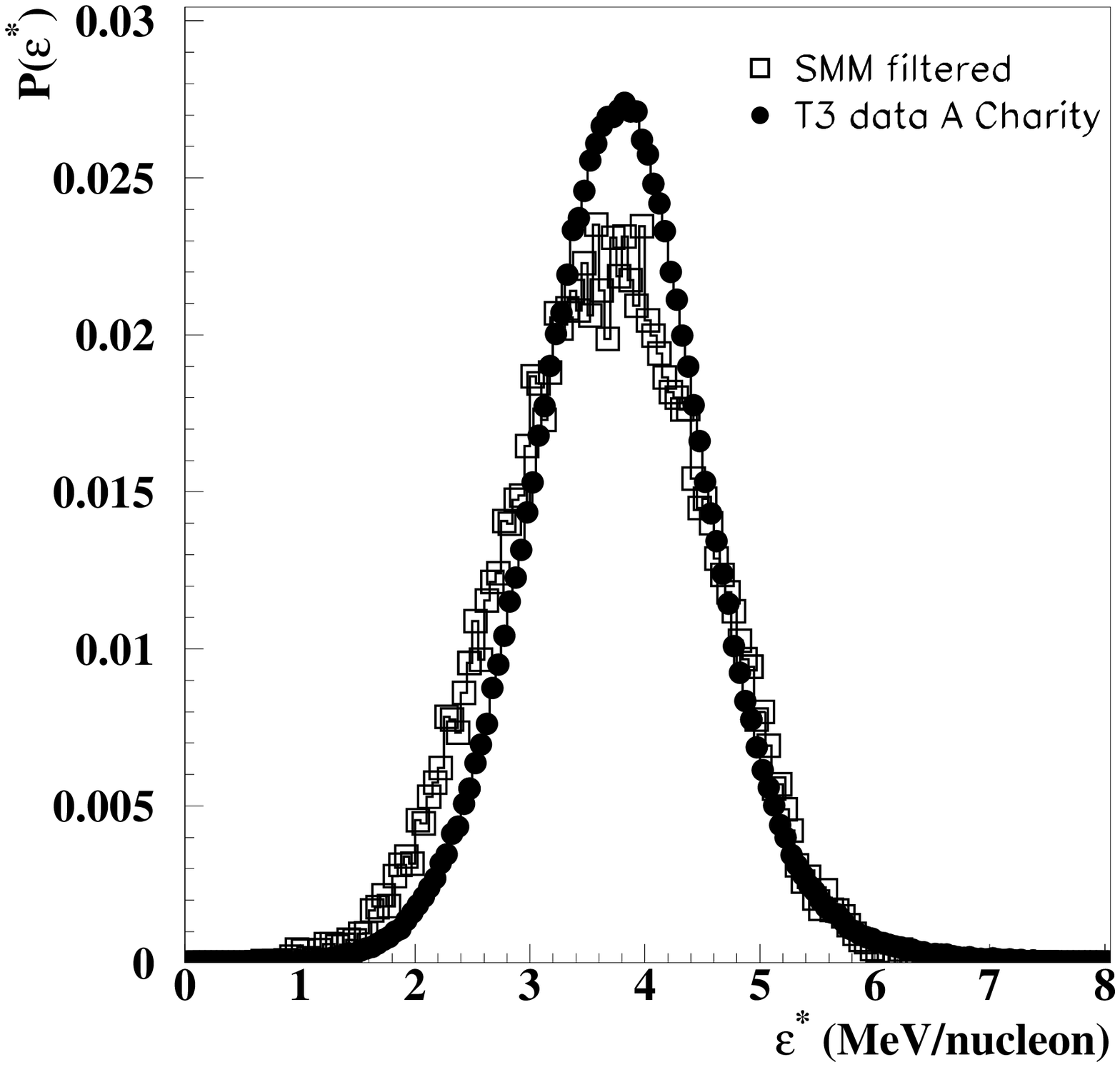}
\caption{\label{back3_3}Source excitation energy for $A_{stable}$ (left) and $A_{Charity}$
(right) mass parameterization. SMM backtraced (open squares) and data 
excitation energy (full points). Region T3.}
\end{figure}

Figures \ref{SMM_back}-\ref{back3_3} present the same comparison as before for 
the SMM calculation. A much better agreement is observed : the residue charge and the total charge 
distributions are nicely 
reproduced and the disagreement on the alpha multiplicity (as well as in the
fragment multiplicity) is less severe than with GEMINI.

It is worth noting that the best agreement with data was found for SMM calculations performed 
with a freeze out volume of $V=2V_0$, and a rotational energy corresponding to an angular momentum
of $35$ $\hbar$ for the QP source. 

%
%

Figures ~\ref{SMM_var1} and ~\ref{back3_3} show the comparison for 
the source variables : they display a good agreement between 
the simulated and the experimental
distributions. The data treated with the  $A_{stable}$ 
mass parameterization show the same shift in the mean value as observed 
with GEMINI.\par

\begin{table}[!h]
\begin{center}
\begin{tabular}{|c|c|c|c|c|}
\hline
 & $Z_s$  & $\epsilon_{stable}^*$ MeV/nucleon &$\epsilon_{Charity}^*$MeV/nucleon & $A_s$\\
\hline
Data& $45.\pm2.$ &$3.4\pm0.5$  & $3.8\pm0.4$&$101.\pm4.$\\
\hline
GEMINI  &$44.\pm2.$ &$3.8\pm0.5$  & $3.8\pm0.5$&$103.\pm4.$\\
\hline
SMM & $45.\pm2.$ &$3.7\pm0.5$ & $3.7\pm0.5$ & $104.\pm3.$\\
\hline
\end{tabular}
\caption{\label{tab4}Data and filtered average values for the source variables obtained from GEMINI
and SMM backtracing for region T3.}
\end{center}
\end{table}
In table ~\ref{tab4} are reported the mean values of the source variables for
the data, for GEMINI and for SMM. Both models give a fair 
agreement with data. One should note that mean values are not sufficient
to quantify the agreement between the experiment and the model.
Indeed, the examination of figures ~\ref{gem_obs} to ~\ref{back3_3}
shows that overall SMM gives a better description of the data than GEMINI.
Therefore we can conclude that for excitation energies above
$3$ MeV/nucleon, the introduction of some degree of multifragmentation 
in the decay of hot nuclei is mandatory in order to reproduce experimental
results.  


\section{Discussion and Conclusions}


We also deduced the average source temperatures from 
both models in order to compare them to the current systematics provided by 
Natowitz et al. in reference \cite{nato}.

In table \ref{tavola} are shown the average temperature and mass values
of the systems for each region.A close comparison of these values for the
region T3 with those obtained for the limiting temperature of reference \cite{nato} shows a 
good agreement : the systems of this region may undergo multifragmentation. This is  
consistent with the conclusions drawn in 
the previous section, i.e. the presence of events to be
described by a multifragmentation scenario.

\begin{table}[!h]
\begin{center}
\begin{tabular}{|c|c|c|c|}
\hline
 & $A_{exp}$  & $T_{Gem}$ MeV &$T_{SMM}$ MeV \\
\hline
T1  &$91.2\pm9.4$   &3.55  & -\\
\hline
T2 & $96.0\pm9.8$ &4.54 & 4.87  \\
\hline
T3 & $101.3\pm9.3$ &5.19& 5.56  \\
\hline
\end{tabular}
\caption{\label{tavola}Mean temperature and mass values for the sources 
of the three regions.}
\end{center}
\end{table}
%
In perspective one should understand why the alpha particle multiplicity is
underestimated by GEMINI. This could be connected to the dependence of the 
level density parameter $\it{a}$ from the excitation 
energy (\cite{leston},\cite{charity}). In the present work, the Lestone
parameterization was used , but there are different formulae
describing this dependence of $\it{a}$ that could be used alternatively 
in the backtracing.

In order to better enhance the differences between the two models, an attempt 
could also be done using the backtracing of dynamical observables , more 
sensitive to the Coulomb effects \cite{napoli}. 
One should however recall that some of the dynamical observables (like
velocities for example) are mass dependent. This addresses the 
problem of the choice of the 
mass parameterization previously evoked. On this subject, we may hope to gain
some useful insight by upgrading in the mass identification of existing $4\pi$ devices like
CHIMERAPS or by new generation projects like FAZIA (Four 4$\pi$ A Z Ion Array)~\cite{fazia}.

\vskip 0.5cm
{\bf Acknowledgements}

This work was supported by the LPC (Laboratoire de Physique Corpusculaires)
Caen and by the AFFDU (Association Fran\c{c}aise des Femmes 
Dipl\^om\'ees des Universit\'es).


\end{document}